\documentclass[amsmath,amssymb,floatfix,showpacs]{revtex4}
\input epsf
\newcommand {\eq}{\begin{equation}}
\newcommand {\qe}{\end{equation}}

\newcommand {\bfq} {{\bf q}}

\newcommand {\bfb} {{\bf b}}

\newcommand {\h}{\frac{1}{2}}

\newcommand {\f}{\lambda}
\newcommand {\ea} {{\it et al.}}

\newcommand {\ar}{a_R}
\newcommand {\ai}{a_I}

\newcommand {\pr}{Phys. Rev. }
\newcommand {\nucp}{Nucl. Phys.}

\usepackage{epsfig}

\begin{document}





\title{Phase Variation of Hadronic Amplitudes}

\vspace*{.5in}

\author{J.-P. Dedonder}
\affiliation{\large Universit\'e Paris Diderot-Paris 7\\
IMNC/Case 7021 B\^atiment Condorcet\\
75205 Paris cedex 13, France}

\author{W. R. Gibbs}

\affiliation{\large New Mexico State University, Las Cruces, 
NM 88003}

\author{Mutazz Nuseirat}

\affiliation{\large King Saud Bin Abdulaziz University, 
P.O.Box 22490, Riyadh 11426, KSA}

\date{\today}

\begin{abstract} 

The phase variation with angle of hadronic amplitudes is studied with a 
view to understanding the underlying physical quantities which control it 
and how well it can be determined in free space.  We find that unitarity 
forces a moderately accurate determination of the phase in standard 
amplitude analyses but that the nucleon-nucleon analyses done to date do 
not give the phase variation needed to achieve a good representation of 
the data in multiple scattering calculations. Models are examined which 
suggest its behavior near forward angles is related to the radii of the 
real and absorptive parts of the interaction. The dependence of this phase 
on model parameters is such that if these radii are modified in the 
nuclear medium (in combination with the change due to the shift in energy 
of the effective amplitude in the medium) then the larger magnitudes of 
the phase needed to fit the data might be attainable, but only for 
negative values of the phase variation parameter.

\end{abstract}


\maketitle

\section{Introduction}

A phenomenological form often used for the nucleon-nucleon scattering 
amplitude is
\eq
f_{ph}(q)=\frac{ik\sigma(1-i\rho)}{4\pi}e^{-\h aq^2},\ \ 
a=\ar+i\ai \label{form}
\qe
where $\sigma$ is the total cross section and $\rho$ is the 
ratio of the real to imaginary part of the forward amplitude. 
This latter can be measured by interference with the coulomb 
amplitude. The parameter $\ar$ can be extracted from the fall 
off of the cross section. At a common value at the beam 
momentum we will consider here (1.75 GeV/c) it is usually taken 
to be 5.6 (GeV/c)$^{-2}$. The parameter $\ai$ has often been 
assumed to be zero for lack of better knowledge.

Although the variation of the phase of the nucleon-nucleon
amplitude with momentum transfer had been considered before \cite{bw,ft}, 
in seminal papers Franco and Yin \cite{fy} found that the 
value of $\ai$ strongly influenced multiple scattering in 
light nuclei and they were able to obtain a much better 
representation of the data if they treated it as a free 
parameter. The values that they found were +10 (GeV/c)$^{-2}$ 
and --15 (GeV/c)$^{-2}$ with either value giving a dramatic 
improvement in the agreement with the data. They considered 
$\alpha$ particle scattering from $^4$He, $^3$He, deuterium 
and $^1$H and found the same improvement for all targets with 
the same values of $\ai$. This, and other studies of multiple 
scattering \cite{lm,lassaut,auger2,usmani,elgogary3,auger}, will be 
discussed in Section \ref{multiple}.

In an attempt to estimate reasonable theoretical values for $\ai$, Ahmad 
and Alvi \cite{ahmad} studied an eikonal \cite{eikonal} approximation 
based on an effective potential and concluded that $\ai$ should be 
expected to have a magnitude of the order of 1 (GeV/c)$^{-2}$ or less, 
apparently in disagreement with the previous work.

However, Ref. \cite{ahmad} considered a potential in which 
the real and imaginary parts had the same spatial 
distribution, a Gaussian form being taken for each component.  
It is generally believed that the nucleon-nucleon interaction 
is more complicated than this and that the ranges to be 
associated with the different parts of the interaction (real 
and imaginary) are significantly different.

A principal aim of this paper is to investigate the effect of 
assuming what is hoped are reasonable estimates for the forms 
of the potential and to understand the relationship of the 
parameter $\ai$ to the geometric structure of the 
interaction.

We will often follow the eikonal method used by Ref. 
\cite{ahmad}. In this approximation one can write
\eq
f_{ei}(q)=\frac{ik}{2\pi}\int d^2b\ e^{i\bfq\cdot\bfb}\Gamma(b)
\label{eikonal}=ik\int_0^{\infty} b db J_0(qb)\Gamma(b)
\qe
where
\eq
\Gamma(b)=1-e^{i\chi(b)}\ \ {\rm and}\ \ 
\chi (b)=-\frac{1}{\hbar v}\int_{ -\infty}^{\infty}V(\sqrt{b^2+z^2})dz,
\label{eikonal2}
\qe
$V(r)$ is a complex potential and $J_0$ is the Bessel function of
zero order.

In order to obtain the forward angle dependence of the phase we can expand 
Eq. \ref{eikonal} in powers of $q$ and equate the coefficients of $q^0$ 
and $q^2$ in the second order expansion of these expressions with the 
expansion of Eq. \ref{form} to find
\eq
a=\frac{1}{2}\frac{\int_0^{\infty} b^3\Gamma(b)db}
{\int_0^{\infty} b\Gamma(b)db}.
\label{adef}\qe

We observe that there is a symmetry that exists in these 
expressions. If the sign of the real part of the potential is 
changed ($V_{Real}\longrightarrow -V_{Real}$) then 
$\Gamma(b)\longrightarrow \Gamma^*(b)$ which means that 
$\rho\longrightarrow -\rho$ and $a_I\longrightarrow -a_I$.

We proceed in the remainder of the paper to investigate the 
origin and dependencies of the phase of the strong scattering 
amplitude. In section II we introduce the relationship of the 
phase to the difference in interaction ranges with a 
schematic model. This model, crude as it is, gives us some 
indication of the interdependencies among the physical 
quantities and the underlying physics involved.

In section III we look at the conditions imposed by unitarity 
on the phase and its determination from phase shift analyses. 
We investigate the accuracy to which the phase can be 
determined in a typical realistic (K$^+$p) case. In section 
IV we show the results of the NN analysis by Arndt \ea 
\cite{arndt}.

In section V we investigate three potential models in order 
to give a somewhat more realistic evaluation of the 
interdependence among the parameters and their variation with 
energy. Section VI treats the question of determining the 
phase variation parameter, $a_I$ from multiple scattering. In 
section VII we draw conclusions from the work and discuss the 
possible changes of the phase in the medium due to nucleonic 
and non-nucleonic mechanisms.

\section{Schematic Delta function model}

As a first orientation, let us consider a schematic model in which 
the strength of the integrand in Eq. \ref{eikonal} is concentrated at 
points in the impact parameter variable, $b$.  In order to see the 
connection between the phase and the radii of the real and imaginary parts 
of the interaction it is useful to define a slightly modified amplitude 
such that the imaginary part is equal to the total cross section

\eq 
F(q)=\frac{4\pi}{k}f(q) \ \ {\rm along\ with}\ \ 
G(b)=4\pi i \Gamma(b)
\qe
so that Eq. \ref{eikonal} becomes 
\eq
F(q)=\frac{1}{2\pi}\int d^2b e^{i\bfq\cdot\bfb}G(b).
\qe

Note that
\eq
F(0)=\sigma(\rho+i).
\label{fzero}
\qe

We now wish to investigate the assumption that the real and imaginary
parts of the strength of $G(b)$ are concentrated in different regions
of impact parameters, hence we consider the simple model 
in which the distribution of the strength of $G(b)$ is expressed by 
$\delta$-functions, i.e., we take
\eq
G(b)=G_R\delta(b-b_R)+iG_I\delta(b-b_I).
\qe
Now
\eq
F(0)=G_R b_R+iG_I b_I=\rho\sigma+i\sigma
\qe
so that the constants $G_R$ and $G_I$ are determined and we can write
\eq
G(b)=\sigma\left[\frac{\rho}{b_R}\delta(b-b_R)+i\frac{1}{b_I}\delta(b-b_I)
\right]
\qe
which, using Eq. \ref{adef}, leads to
\eq
a=\ar+i\ai=\frac{\rho b_R^2+ib_I^2}{2(\rho+i)}
\label{a}
\qe
or
\eq
\ar=\frac{\h(b_I^2+\rho^2b_R^2)}{1+\rho^2},\ \ 
\ai=\h\rho\left(\frac{b_I^2-b_R^2}{1+\rho^2}\right).
\qe
Thus we see that $a_I$ is directly related to the difference in the radii 
of the real and imaginary parts.

If we assume that $b_R$ is to be associated with one pion exchange, 
dominant in this region of low momentum transfer \cite{gl}, and the 
absorption radius with the two-pion exchange range \cite{benoit}, or about 
half as big, then, with a one-pion-exchange range of 1.4 fm we have,
\eq \h(b^2_I-b^2_R)=
\h(0.7^2-1.4^2)\ {\rm fm}^2=-0.75\ {\rm fm}^2 \approx -19.3 
\ {\rm (GeV/c)}^{-2}
\qe
so that $a_I$ is roughly proportional to $\rho$ (for small values) with a 
relatively large coefficient. If we were to take the two radii to be equal 
(as was done in Ref. \cite{ahmad}) the coefficient would be zero. We also 
see a strong correlation between $a_I$ and $\rho$.  We will see later that 
these general features are present in more realistic potential models. The 
relation between the phase and impact parameters has been discussed before 
\cite{kundrat}, although at much higher energies (P$_{\rm Lab}\ \ge 100$\ 
GeV/c).

\section{Unitary Constraints}

We now investigate to what extent the fact that physical 
amplitudes have an expression in partial waves with 
coefficients which satisfy unitary constraints restricts the 
phase parameter $a_I$. We first consider the model amplitude 
in Eq. \ref{form} and then treat the general case.

\subsection{Unitary Limits for a Gaussian Amplitude}

In the form used in Eq. \ref{form} there are limits on the 
values $\ar$ and $\ai$ can take from a unitary expansion of 
the amplitude, assuming the amplitude represents scattering 
of a zero-spin projectile on a zero-spin target. We can write 
\eq \frac{ik\sigma (1-i\rho)}{4\pi}e^{-\h aq^2} 
=\frac{ik\sigma (1-i\rho)}{4\pi} e^{-a 
k^2}\sum_{\ell=0}^\infty i^{\ell}(2\ell+1)j_{\ell}(-ia 
k^2)P_{\ell}(x) =\frac{1}{2ik}\sum_{\ell=0}^\infty 
(2\ell+1)(S_{\ell}-1) P_{\ell}(x) 
\qe 
where 
\eq 
q^2=2k^2(1-x) 
\qe 
and $x=\cos\theta$. Thus, we can identify 
\eq 
S_{\ell}=1-\frac{k^2\sigma(1-i\rho)}{2\pi}e^{-ak^2}i^{\ell} 
j_{\ell}(-iak^2). \label{sl} 
\qe 

The fact that the absolute values of $S_{\ell}$ cannot exceed unity leads 
to the condition 
\eq 1 \ge |S_{\ell}|^2=1-2\mu(R_{\ell}+\rho 
I_{\ell}) +\mu^2(1+\rho^2)(R_{\ell}^2+I_{\ell}^2)\label{cond} 
\qe 
where 
\eq \mu=\frac{k^2\sigma}{2\pi} 
\qe 
and 
\eq 
R_{\ell}=Re[i^{\ell}e^{-ak^2}j_{\ell}(-iak^2)],\ \ \ 
I_{\ell}=Im[i^{\ell}e^{-ak^2}j_{\ell}(-iak^2)] .
\qe

For $a$ real ($=\ar$), so that $I_{\ell}=0$, we can write
the condition \ref{cond} as
\eq
\mu R_{\ell}\le \frac{2}{1+\rho^2} \label{cond1}
\qe
where $R_{\ell}$ is the real quantity $i^{\ell} e^{-\ar 
k^2}j_{\ell}(-i\ar k^2)$. While this condition must hold for 
all partial waves, numerical studies indicate that the s-wave 
unitarity is the most likely to be violated. For this case we 
have the result
\eq
1-e^{-2\ar k^2}\le\frac{8\pi \ar}{\sigma(1+\rho^2)},
\qe
which can be regarded as a constraint on $a_R$ or $\rho$.

This condition is similar to, but stronger than, the constraint arising 
from the requirement that the integrated elastic cross section is less 
than or equal to the total cross section which is

\eq
1-e^{-4\ar k^2}\le\frac{16\pi \ar}{\sigma(1+\rho^2)}.
\qe

With values of $\ar$ satisfying the condition given by Eq. \ref{cond1} we
included finite values of $\ai$ and studied numerically the 
resulting values of $|S_{\ell}|^2$. It was found that
for values corresponding to large partial waves (where
$|S_{\ell}|$ is nearly unity in any case), unitarity
was violated to some (often small) extent. 

We can see in the following that, in the limit of large $\ell$, unitarity 
in some partial waves must be violated. Using the limit for large $\ell$ 
for the spherical Bessel function
\eq
i^{\ell}e^{-ak^2}j_{\ell}(-iak^2)\longrightarrow
\sqrt{\frac{e}{2}}i^{\ell}e^{-ak^2}\frac{(-iak^2e)^{\ell}}
{(2\ell+1)^{\ell+1}}
=\sqrt{\frac{e}{2}}e^{-ak^2}(a_Rk^2e)^{\ell}\frac{(1+ia_I/a_R)^{\ell}}
{(2\ell+1)^{\ell+1}}
\longrightarrow 
\sqrt{\frac{e}{2}}\frac{(a_Rk^2e)^{\ell}e^{-a_Rk^2}}
{(2\ell+1)^{\ell+1}}e^{i\chi_{\ell}}
\qe
where
\eq
\chi_{\ell}\equiv -a_Ik^2+\ell\frac{a_I}{a_R}.
\qe
Since $R_{\ell}$ and $I_{\ell}$ are going to zero with increasing $\ell$
we can drop the last term in Eq. \ref{cond} to get
\eq
1\ge 1-\sqrt{2e}\mu \frac{(a_Rk^2e)^{\ell}e^{-a_Rk^2}}
{(2\ell+1)^{\ell+1}}
[\cos\chi_{\ell}+\rho\sin\chi_{\ell}].
\qe
Since for some values of $\ell$ the quantity in brackets must be
negative, we see that the condition will be violated for some
S-matrix elements.

One important caveat is that this proof holds only
for the scattering of two spin-zero particles since only in that
case can the amplitude be written in the unitarity form we have taken.
The Gaussian expression for the amplitude is often used for a 
spin-averaged amplitude which is not expressible in this form.

Thus, for the strict respect of unitarity, the form of Eq. 
\ref{form} requires that $\ai\equiv 0$ and that condition 
\ref{cond1} holds. Of course, one could modify the values of 
$S_{\ell}$ in any partial wave in which unitarity did not hold 
(i.e. so that Eq. \ref{sl} is no longer true) but the manner of 
carrying out this correction is non-unique and the functional form 
is (perhaps only slightly) different from Eq. \ref{form}. We 
discuss this problem from a more general point of view in the next 
section.

\begin{figure}[htb] 
\epsfig{file=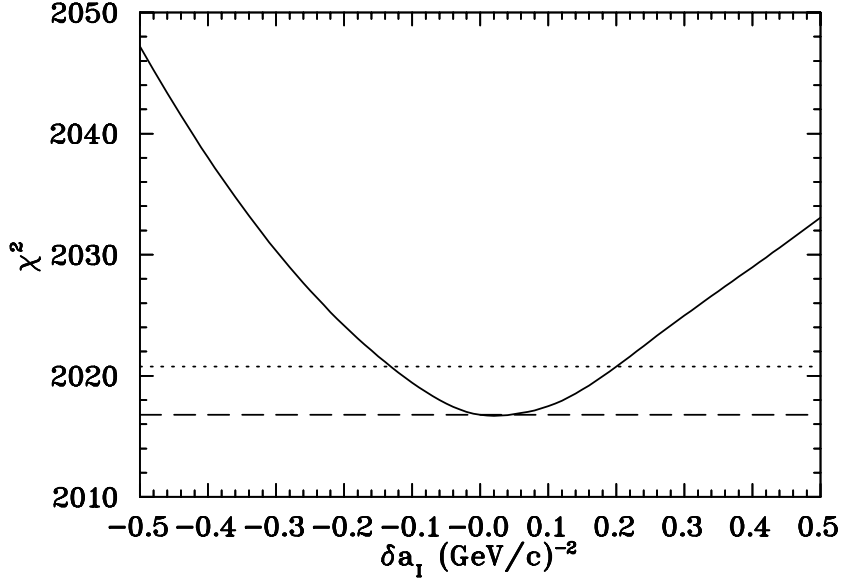,angle=90,height=3.in} 
\caption{Values of $\chi^2_d$ as a function of the deviation of the phase 
from the original one. The dashed line indicates the minimum $\chi^2$ and 
the dotted line is drawn at a value 4 units larger} \label{gamm5pts} 
\end{figure}

\begin{figure}[htb] 
\epsfig{file=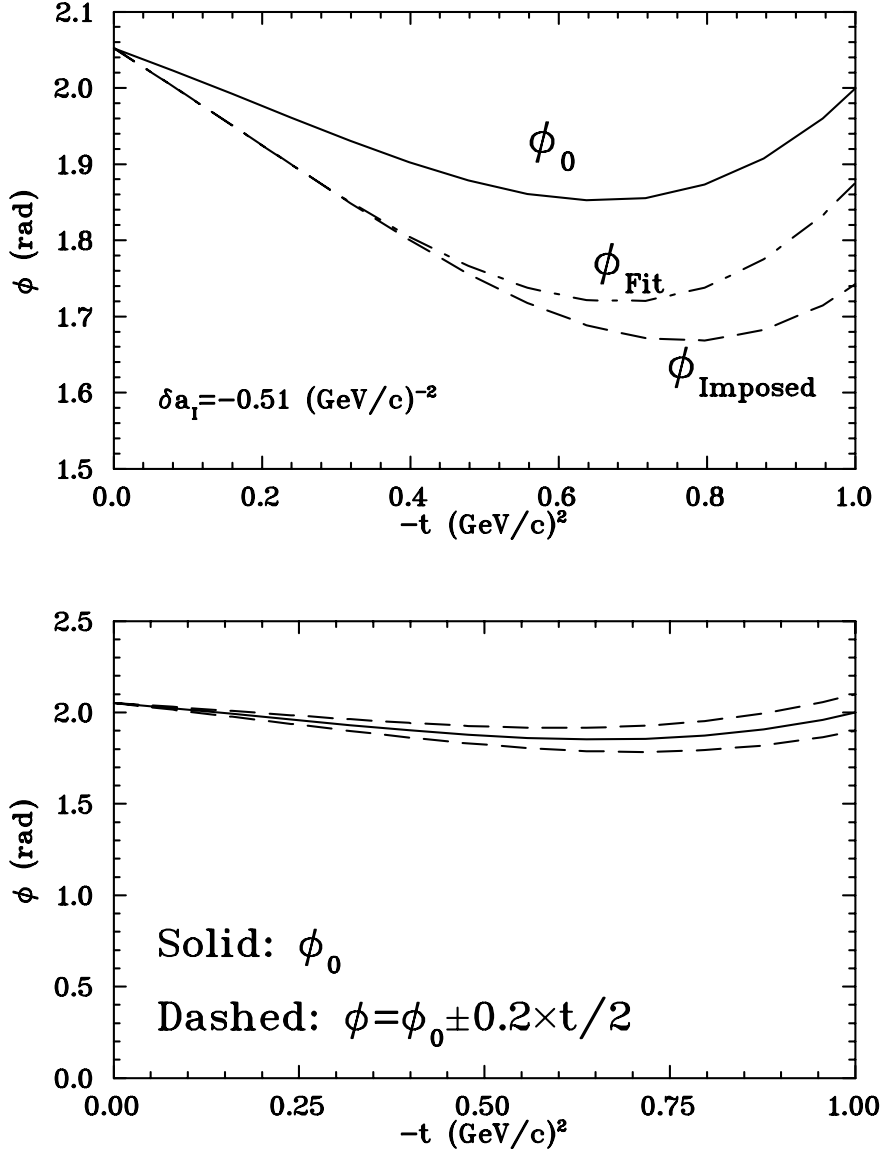,height=6.in} 
\caption{Comparison of phases of K$^+$p scattering. Upper panel: The solid 
curve gives the phase from the original amplitude and the dashed curve the 
result of modifying it with an additional phase with $\delta\ai=-0.51$ 
(GeV/c)$^{-2}$. The dash-dot curve gives the phase of the amplitude 
resulting 
to the best fit to the experimental data plus the phase data out to 0.32 
(GeV/c)$^2$. The phases match out to about 0.36 (GeV/c)$^2$. The case 
chosen for this comparison is the worst case (has the largest value of 
$\chi^2$ from the phase ``data''). Lower panel: Original phase and limits 
corresponding to about two standard deviations in the classical estimate. 
The values that correspond to this point on Fig. \ref{gamm5pts} are about 
$\pm$ 0.2 (GeV/c)$^{-2}$.} \label{ph}
\end{figure}

\subsection{Unitary Limits in Amplitude Analyses}

A similar technique to that used in the previous section can be applied to 
a more general amplitude. We are particularly interested here in the use 
of a partial-wave expansion to represent experimental data
and the question of how well the phase is determined.

To this end, we first consider the case, again, for spin-zero 
on spin-zero scattering, where one attempts to change the 
phase of the amplitude by an arbitrary function, 
$\phi(\theta)$ to obtain an amplitude with a different phase.

\eq
\widetilde{f}(\theta)=e^{i\phi(\theta)}f(\theta)
\label{ftilde}\qe

It has often been assumed that such a phase would be 
undetectable in elastic scattering since it does not affect 
the measurable cross section. It does, however, modify 
unitarity in a manner similar to that seen in the previous
section. 

We will take the form of $\phi(\theta)$ to be linear in $t$, 
i.e. 
\eq
\phi(\theta)=-\h\delta \ai t=\h\delta\ai 2 k^2(1-\cos\theta)\label{phaseform}
\qe
where $\nu=\delta\ai  k^2$. We can expand both 
amplitudes in Eq. \ref{ftilde} in Legendre series to find
\eq
f(\theta)=\sum_{\ell}f_{\ell}P_{\ell}(\cos\theta)
\ \ {\rm and}\ \ \widetilde{f}(\theta)=
\sum_L\widetilde{f}_L P_L(\cos\theta)\label{partialwave}
\qe
where 
\eq
f_{\ell}=\frac{S_{\ell}-1}{2ik}\ \ {\rm and} \ \ 
\widetilde{f}_L=\frac{\widetilde{S}_L-1}
{2ik}\label{smat}.
\qe
Using Bauer's series,
\eq
e^{-i\nu x}=\sum_{\lambda}i^{\lambda}(2\lambda+1)
P_{\lambda}(\cos\theta) j_{\lambda}(-\nu),
\qe
and Eqs. \ref{ftilde}, \ref{phaseform} and 
\ref{partialwave}, we can express the coefficients in the 
expansion of $\widetilde{f}(\theta)$ as
\eq
\widetilde{f}_L=e^{i\nu}\sum_{\lambda,\ell}
(2\lambda+1)i^{\lambda}j_{\lambda}(-\nu)f_{\ell}
\left[ C_{L,\lambda,\ell}^{0,0,0}\right]^2\label{fltilde}
\qe
where $C_{L_1,L_2,L_3}^{M_1,M_2,M_3}$ is a Clebsch-Gordon
coefficient. Solving for the S-matrix element $\widetilde{S}_L$
from Eqs. \ref{smat} and \ref{fltilde} one can check if
$|\widetilde{S}_L|^2\le 1$. We find, with numerical 
studies, that unitarity is always violated in some partial wave.

How to judge the seriousness of a violation of unitarity is 
perhaps not obvious.  One way to do so is to correct the 
violation and see what difference the change makes in the cross 
section and other observables derived from the new amplitude.  
This leads then to constraints due to the data.  Rather 
than make {\it ad hoc} changes to the S-matrix elements, it is 
preferable to perform a search fitting the data to make the 
decision about how the parameters are to be changed to preserve 
unitarity, give the desired phase and, at the same time, give 
the best fit to the data in the sense of a lowest $\chi^2$.

We have implemented this idea in a fitting program for K$^+$p scattering 
\cite{ga}, a system for which the data are relatively good. The isospin 
structure is the same as in nucleon-nucleon scattering but the spin 
structure is less complex.  For K$^+$p scattering there is no 
one-pion-exchange (OPE) contribution as there is in the nucleon-nucleon 
case. It has been suggested that OPE will help to determine the phase of 
the amplitude for NN scattering \cite{bystricky} but in the present test 
we have only the unitarity constraints to determine the phase.

We now study how well the phase can be determined in the 
process of finding a fit of a unitary form to the data.  For 
K$^+$p scattering we can write the amplitude as 
$F(\theta)+\sigma\cdot{\bf n}G(\theta)$ where ${\bf n}$\ is a 
unit vector perpendicular to the scattering plane. In the 
course of this work a minimum lower than those found in Ref. 
\cite{ga} was observed. The original best $\chi^2$ found was 
2031.05 while the new one is at 2016.77. While the difference 
in $\chi^2$ per data point is very small, as is the change in 
the phase shift parameters, the difference in 
$\chi^2$ is important for the calculation of error estimates.

The procedure used is first to calculate the ``natural'' phase, $\phi_0$, 
obtained from the original fit to the experimental data. We then calculate 
a new phase, the ``imposed'' phase $\phi=\phi_0-\delta\ai t$. In the 
fitting procedure, points from artificial data are included which consist 
of values of the phase to be imposed at a chosen single energy over a 
restricted range of $t$.  The new phase ``data'' are included in the 
non-spin-flip amplitude, $F(\theta)$, only, the spin-flip amplitude being 
left free to have whatever phase the fit prefers. This additional phase 
``data'' is taken to have very small errors to force the desired phase. A 
constant error (0.001 rad) is taken at each of 5 phase data points. In the 
modified minimization process there are two contributions to $\chi^2$:

\eq
\chi^2=\chi^2_d+\chi^2_{\phi}
\qe
where $\chi^2_d$ is the part of the $\chi^2$ from the 
experimental data points and $\chi^2_{\phi}$ is that coming 
from the phase points.

The total $\chi^2$ (including the phase data) is minimized 
but the part of $\chi^2$ of principal interest is that due 
to the experimental data points, $\chi_d$.  It is found, as 
anticipated, that the $\chi^2_d$ from the true data increases 
as the imposed phase is chosen farther from the original 
phase, $\phi_0$.

The result of $\chi^2_d$ for fitting the forward phase up to 
--t=0.32 (GeV/c)$^2$ at a beam momentum of 1.3 GeV/c is shown in 
Fig. \ref{gamm5pts}. An increase in $\chi^2_d$ of 4 gives an 
estimate of the uncertainty of $\pm0.2$\ (GeV/c)$^{-2}$.

\begin{figure}[htb]
\epsfig{file=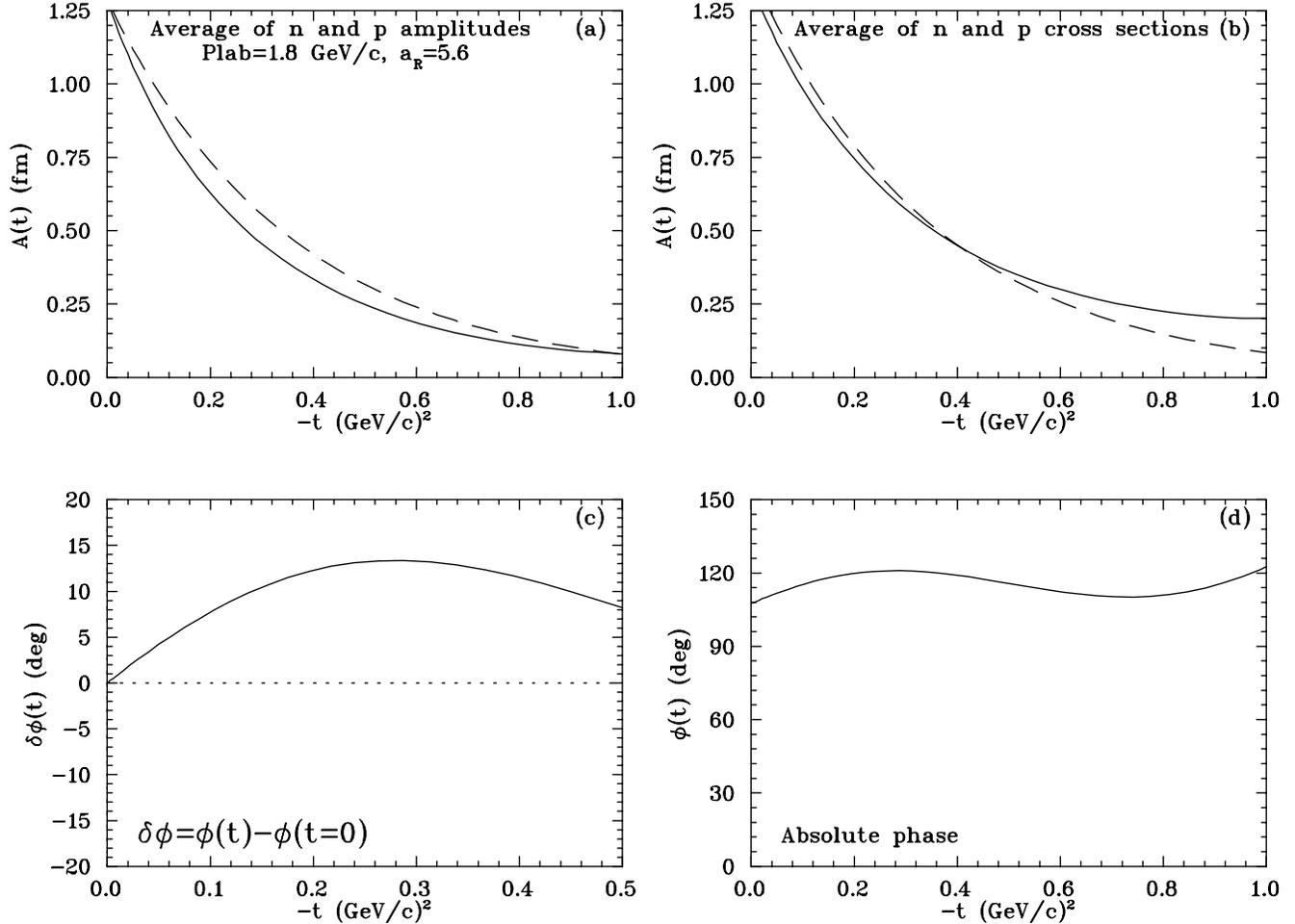,angle=90,height=5.in}
\caption{Properties of the NN amplitudes at 1.8 GeV/c. Panel (a) 
shows the result of the absolute value of the averaged neutron 
and proton amplitudes. In panel (b) the solid curve shows the 
result of averaging the cross sections and taking the square 
root. In both (a) and (b) the dash curve shows a plot of the 
Gaussian approximation with $\ar=5.6\ (GeV/c)^{-2}$. Panel (c) 
shows the phase relative to the phase at $t=0$ (solid curve). 
Panel (d) shows the absolute phase over an extended range of 
$t$.}
\label{phasearndt} 
\end{figure}

\begin{figure}[htb]
\epsfig{file=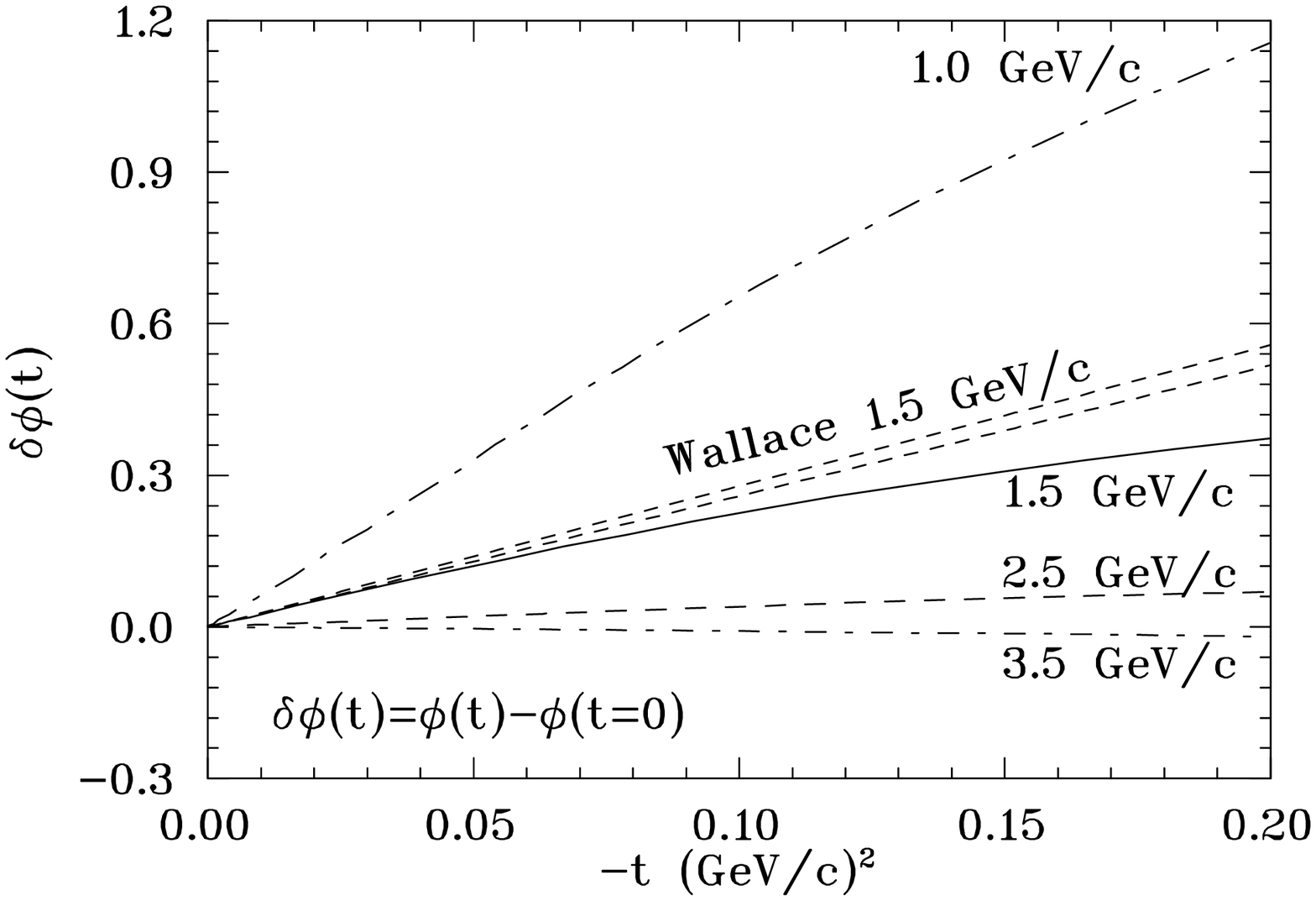,angle=90,height=5.in} 
\caption{Dependence of the phase on momentum transfer. The 
curves labeled with a momentum (P$_{Lab}$) are from Arndt et al. 
\cite{arndt}. The two dashed curves labeled ``Wallace'' are the 
phases from neutron and proton spin averaged amplitudes from 
Ref. \cite{wallace2}. }
\label{phasearndt1} \end{figure}

In the upper part of Fig. \ref{ph} one sees that the imposed 
phase is well fit up to 0.4 (GeV/c)$^2$. Attempting to fit to 
a larger range of $t$ did not lead to a better fit to the 
desired phase but instead the $\chi^2$ due to the phase 
portion became larger. It appears that the limit has been 
reached where a modification of the phase linear in $t$ can 
easily be used. The lower part of Fig. \ref{ph} shows the 
limits of $\pm$ 0.2 (GeV/c)$^{-2}$, an uncertainty of around 
$\pm$ 5\%. Thus, the phase is reasonably well determined. One 
might believe that the phase of the nucleon-nucleon amplitude 
is better determined because of the use of OPE in the higher 
partial waves \cite{bystricky}.

\section{Nucleon-nucleon Amplitude}

With some confidence that a phase-shift fit to data provides 
a reasonably reliable determination of the absolute phase, we 
now turn to the nucleon-nucleon phase obtained from the 
recent fit by Arndt et al. \cite{arndt}.  The spin average is 
given by

\eq
\bar{M}=\h (<++|M|++>+<+-|M|+->)
=\frac{M_{11}+\h M_{ss}+\h M_{00}}{2}.
\qe

The amplitude must also be averaged over the neutron
and proton so that the averaged amplitude becomes
\eq
A=\frac{3}{4}\bar{M}(I=1)+\frac{1}{4}\bar{M}(I=0)
\qe

In Fig. \ref{phasearndt} are shown several properties of the 
Arndt et al. \cite{arndt} amplitudes at 1.8 GeV/c.  The value of 
$\ar$ can be extracted by averaging the amplitudes or the cross 
sections. The top panels show the result of the two methods 
compared with the exponential form given in Eq. \ref{form}. The 
panel (c) shows the variation of the phase from the forward 
value. It is seen that a linear approximation in $t$ is 
reasonable up to $-t$ of about 0.2 (GeV/c)$^2$. 
Panel (d) shows the variation of the total phase.

Figure \ref{phasearndt1} shows the variation of the phase for 
several incident momenta as extracted from Arndt et al. 
\cite{arndt} for various beam momenta. Also shown are values from 
Wallace \cite{wallace2} for spin-independent neutron and proton 
amplitudes. Wallace's values were taken from earlier fits by 
Arndt's group at lower energies. It is seen that for all cases 
except for the very highest beam momenta the phase increases with 
$q^2=-t$ so that $a_I<0$.

\section{Potential Models}

In this section we consider potential models for the interaction in an 
attempt to relate the phase to underlying physical parameters. For the 
Gaussian potential in the eikonal approximation we can make considerable 
progress analytically.  We also consider the exact numerical solution for 
this form of potential as well as that of exponential and Woods-Saxon 
potential forms.

\subsection{Gaussian Potentials}

In this section we assume a potential expressed as
\eq
V(r)=V_Re^{-r^2/r_R^2}+iV_Ie^{-r^2/r_I^2}.
\qe
The integral on z in Eq. \ref{eikonal2} can be easily done to give
\eq
i\chi(b)=-\left(\alpha e^{-b^2/r_R^2}+\beta e^{-b^2/r_I^2}\right)
\qe
where $\alpha=i\sqrt{\pi}r_RV_R/v=i\alpha'$ is purely imaginary and
$\beta=-\sqrt{\pi}r_IV_I/v$ is real and positive (since $V_I$ must
be negative).

Now the eikonal expression is
\eq
f(q)=-ik\int_0^{\infty} b db J_0(bq)\sum_{n=1}^{\infty} \frac{(-1)^n}{n!}
\left(\alpha e^{-b^2/r_R^2}+\beta e^{-b^2/r_I^2}\right)^n
\qe
\eq
=-ik\int_0^{\infty} b db J_0(bq)\sum_{n=1}^{\infty} \frac{(-1)^n}{n!}
\sum_{m=0}^n \left( \begin{array}{c} n\\ m\end{array}\right)
\alpha^{n-m}\beta^me^{-(n-m+\eta m)b^2/r_R^2}
\label{expand1}
\qe
where $\eta\equiv r_R^2/r_I^2$. Following the development in 
Appendix A the full expansion of the forward amplitude in powers of 
$\alpha$ is given by
\eq
f(0)=\frac{-ikr_I^2}{2}\left[ 
\f(0,1,\beta)+\sum_{\ell=1}^{\infty}
\frac{(-\alpha)^{\ell}}{\ell !}\f_0 
(\frac{\ell}{\eta},1,\beta)
\right].\label{sum0}
\qe

We have introduced a generalization of the incomplete Gamma function
with $u\ge 0$,
\eq
\f(u,k,\beta)=\sum_{n=1}^{\infty}\frac{(-\beta)^n}
{n!(u+n)^k}.\label{lamdef}
\qe
This function can be computed from its expansion over a large part of
its range. Some of its properties, including an 
asymptotic expansion for large final argument, are discussed in Appendix 
B. Another function, useful when the first argument of $\lambda$ is 
non-zero, is
\eq
\f_0(u,k,\beta)=\sum_{n=0}^{\infty}\frac{(-\beta)^n}
{n!(u+n)^k}=u^{-k}+\f(u,k,\beta).
\qe

Using Eq. \ref{sum0} for $f(0)$ we can see that (to lowest 
order in $\alpha$)
\eq
\rho=\alpha' \frac{\eta+\f(\frac{1}{\eta},1,\beta)}
{\f (0,1,\beta)}
\qe
where $\f (0,1,\beta)<0$. The proportionality of $\rho$ to $\alpha'$
shows that it is strongly influenced by the real part of the potential.

The total cross section, in lowest order in $\alpha$, is given by 
\eq
\sigma=2\pi r_I^2[g_1+\ln (\beta)+E_1(\beta)] \label{totalcross}
\qe
where $g_1$ is Euler's constant. Since, for the range of values of $\beta$
used here, $E_1(\beta)$ is small, the dependence of $\sigma$ on $\beta$ is
small so that this equation provides a strong constraint on $r_I$. The 
lowest order correction in $\alpha$ is $\alpha^2$.
This is an exact 
representation of the simple eikonal approximation (i.e. 
without corrections given by Wallace \cite{eikonal}) for a 
purely absorptive potential.

\begin{figure}[htb]
\epsfig{file=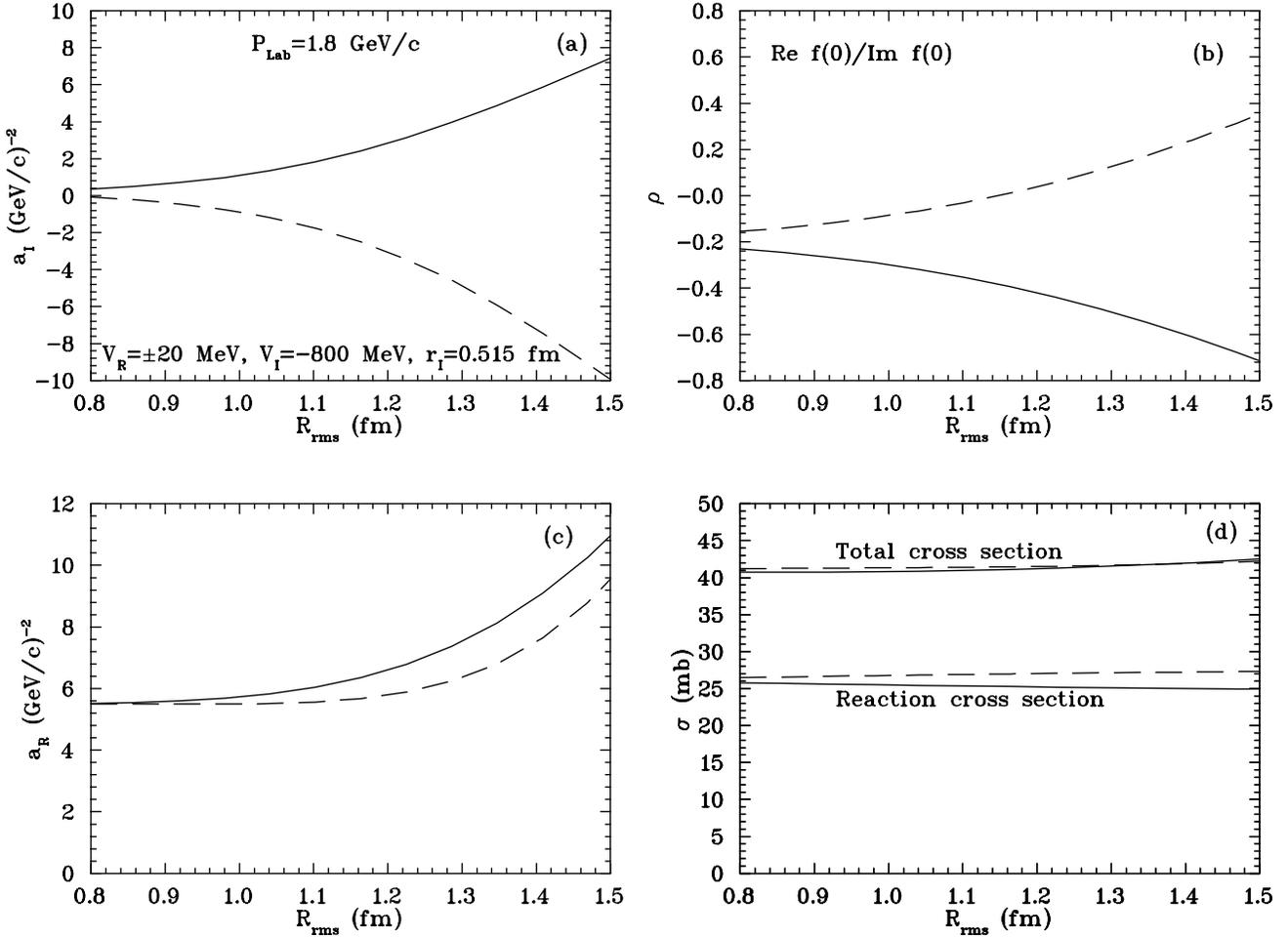,angle=90,height=5.in}
\caption{Dependence of several observables on the r.m.s. radius of 
the real Gaussian potential well for both signs of its strength. 
The solid (dashed) curve shows the result for the positive 
(negative) potential. Panel (a) shows that the phase parameter has 
a very strong dependence on the radius of the real potential. In 
panel (b) it is seen that the ratio of real to imaginary part of 
the forward amplitude also has a fairly strong dependence while the 
Gaussian parameter controlling the fall-off of the absolute value 
and the total and reaction cross sections have little or no 
dependence on this parameter.}
\label{gamrms}
\end{figure}

Equation \ref{a14} can be expressed as
\eq
f(q)=\frac{-ikr_I^2}{2}
\sum_{j=0}^{\infty}
\frac{1}{j!}\left[\f (0,j+1,\beta)+\sum_{\ell=1}^{\infty}
\frac{(-\alpha)^{\ell}}{\ell !} \f_0(\frac{\ell}{\eta},j+1,\beta)\right]
\left(\frac{-q^2r_I^2}{4}\right)^j. 
\label{43}
\qe

We can write the ratio to the forward amplitude (Eq. \ref{sum0}) as
\eq \frac{f(q)}{f(0)}=\frac{\sum_{j=0}^{\infty}
\frac{1}{j!}\left[\f (0,j+1,\beta)+\sum_{\ell=1}^{\infty}
\frac{(-\alpha)^{\ell}}{\ell !} \f_0(\frac{\ell}{\eta},j+1,\beta)\right]
\left(\frac{-q^2r_I^2}{4}\right)^j}{\f(0,1,\beta)+\sum_{\ell=1}^{\infty}
\frac{(-\alpha)^{\ell}}{\ell !}\f_0 
(\frac{\ell}{\eta},1,\beta)}
\qe
so that, retaining only the first order contribution in 
$\alpha'$ and $q^2$, we have,
\eq \frac{f(q)}{f(0)}
=1-\frac{r_I^2q^2}{4}\frac{\f(0,2,\beta)}{\f(0,1,\beta)}
\left[
1+i\alpha'\frac{\f_0(\frac{1}{\eta},1,\beta)}{\f(0,1,\beta)}
-i\alpha'\frac{\f_0(\frac{1}{\eta},2,\beta)}{\f(0,2,\beta)}
\right].
\qe
From this expression we can identify 
\eq
\ar+i\ai=
\frac{r_I^2}{2}\frac{\f(0,2,\beta)}{\f(0,1,\beta)}
\left\{
1+i\alpha'\left[\frac{\f_0(\frac{1}{\eta},1,\beta)}{\f(0,1,\beta)}
-\frac{\f_0(\frac{1}{\eta},2,\beta)}{\f(0,2,\beta)}\right]
\right\}
\qe

\begin{figure}[htb]
\epsfig{file=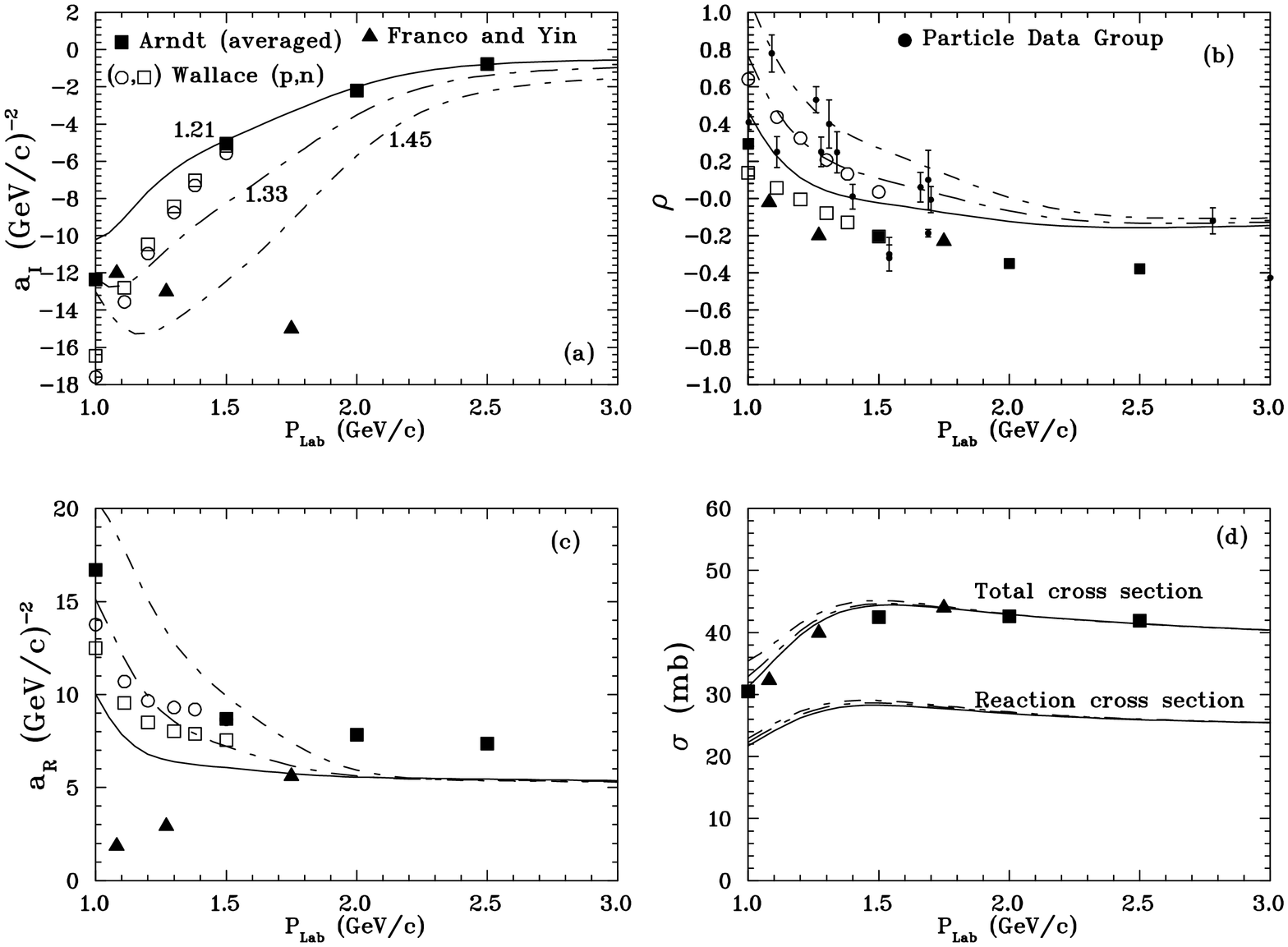,angle=90,height=5in}
\caption{Dependence of several parameters on the radius of 
the real potential well for the exponential potential. Panel 
(a) shows that the phase parameter has a very strong 
dependence on the radius of the real potential. The curves 
are labeled by the real radius (in fm). These labels identify 
the curves for the remaining panels as well. In panel (b) it 
is seen that the ratio of real to imaginary part of the 
forward amplitude also has a fairly strong dependence while 
(panels c and d) the parameter controlling the fall-off of 
the absolute value and the total and reaction cross sections 
have little or no dependence on this parameter at the higher 
energies. The solid 
triangles denote the values used by Franco and Yin \cite{fy}. 
The open circles and squares where taken from Wallace 
\cite{wallace2}. The solid squares were extracted in the 
present work from the fit of Arndt et al. \cite{arndt} and 
were fitted to get the solid curve. The solid dots with 
errors in panel (b) were taken for the Particle Data Group 
\cite{pdg} with several points at low laboratory momentum 
with negative values excluded.}
\label{solexpv}
\end{figure}

Thus, in this linear approximation in $\alpha=i\alpha'$, we see that $\ar$
is relatively stable under variations in the radii. However,
the two ratios multiplying $\alpha'$ are very similar for equal
radii so that $\ai$ varies rapidly for small variations in
the radii of the real and imaginary potentials.

Figure \ref{gamrms} shows several observables obtained from 
exact solutions for the Gaussian potential of as a function 
of the radius of the real part of the potential for $V_R=\pm 
20$ MeV. The sign change of $\rho$ and $a_I$ with sign change 
of the real potential noted in the introduction in the 
eikonal prescription is approximately reproduced in the 
exact calculations.

\subsection{Exponential Potential}

In this section we look at potentials of the form
\eq
V(r)=W_R e^{-r/c_R}+iW_Ie^{-r/c_I}.
\qe

We take both $W_R$ and $W_I$ to be functions of the incident 
momentum in order to fit the data. We have used the phase 
shift analysis of Arndt et al. \cite{arndt} to calculate the 
spin-isospin average of the amplitude and used these 
amplitudes as a guide to fitting the potentials so we are 
using a spin-zero on spin-zero calculation to fit spin 
averaged data. We also used the older results of Wallace 
\cite{wallace2} as well as the values of $\rho$ from the 
Particle Data Group \cite{pdg}. Figure \ref{solexpv} shows 
the results.

We see (solid line, direct fit) that the phase parameter 
$a_I$ is negative at low incident momentum and appears to be 
nearing zero at higher momenta. The parameter $\rho$ shows a 
similar behavior with opposite sign as expected from the 
simple $\delta$-function model of Section II.  The parameter 
$a_R$ rises slightly at low momenta and the total and 
reaction cross sections are relatively flat although the 
reaction cross section goes to zero as the momentum is 
reduced below 1 GeV/c.

Also shown in Fig. \ref{solexpv} are curves for changes from 
the fitted r.m.s. value for the radius of the real potential 
of 1.21 fm to 1.33 fm (dot-long dash) and 1.45 fm (dot-short 
dash) while holding all other parameters fixed.. It is seen 
that there is a significant change in $a_I$ and a 
corresponding change in $\rho$ as well.  There is also a more 
moderate (except at low momentum) change in $a_R$.
At 2.0 GeV/c $a_I$ goes from --2.01 to --3.52 to --5.70 (GeV/c)$^{-2}$
as the percentage change in radius goes from 0 to 10\% to 20\%.

\begin{figure}[htb]
\epsfig{file=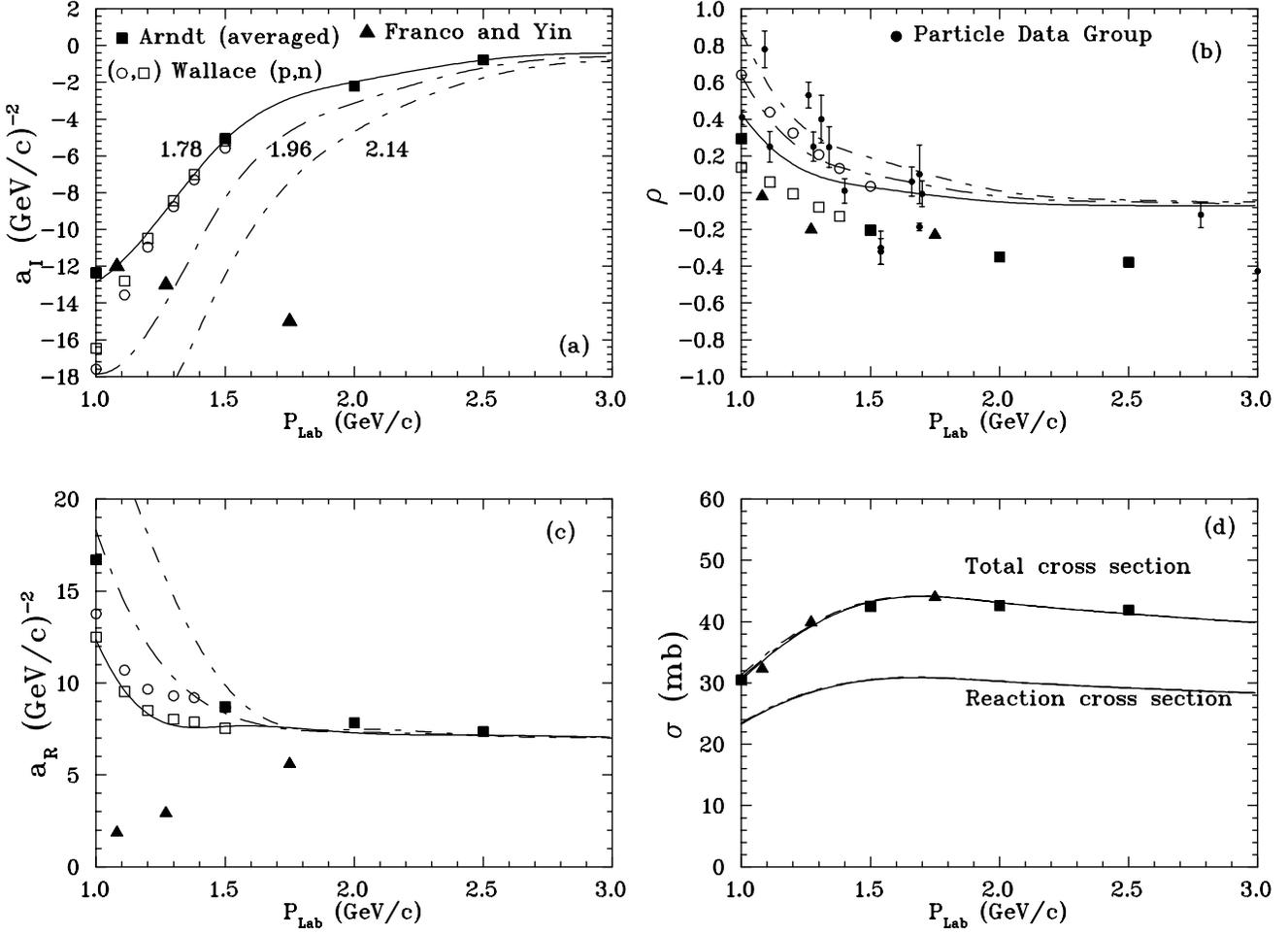,angle=90,height=5in}
\caption{Dependence of several parameters on the radius of 
the real potential well for the Woods-Saxon potential. Panel 
(a) again shows that the phase parameter has a very strong 
dependence on the radius of the real potential. The curves 
are labeled in panel (a) by the rms value of the radius of 
the real Woods-Saxon potential (in fm). These labels identify the curves 
for the remaining panels as well. The curves and symbols have the
same meaning as in Fig. \ref{solexpv}.
}
\label{solws1}
\end{figure}

\subsection{Woods-Saxon Potential}

In this section we look at potentials of the form
\eq
V(r)=\frac{U_R}{1+ e^{(r-d_R)/p_R}}+i\frac{U_I}{1+e^{(r-d_I)/p_I}}.
\qe
We do not believe that this form provides a realistic representation of
the distribution of the strength of the interaction for the 
nucleon-nucleon system but we include it to show that the variation with 
percent increase in the radius of the real potential is about the same as 
the Gaussian or exponential potential.

Figure \ref{solws1} shows results for a fit with this 
potential. We see that while the basic fit has a larger 
r.m.s. radius for the real potential than for the exponential 
potential, the change with percentage change in radius is 
qualitatively very similar. At 2.0 GeV/c $a_I$ goes from 
--1.96 to --3.13 to --4.68 (GeV/c)$^{-2}$ as the percentage 
change in radius goes from 0 to 10\% to 20\%.

\section{Multiple Scattering Results\label{multiple}}

Since multiple scattering of the projectile on a nuclear 
target depends on the phase of the amplitude it may be 
possible to measure the phase (including the phase variation)
in this manner. Of course, one must be aware that what is 
obtained is the effective value in a nuclear medium.

\subsection{Light Nuclei}

Franco and Yin \cite{fy} considered the amplitude as defined 
in the introduction where it was assumed that only the 
non-spin-flip amplitude was needed for the elastic scattering 
on light nuclei that they considered and the variation of 
phase was relative to the phase at zero degrees. We follow 
this same definition. They pointed out that if the ratio of 
real to imaginary part of the amplitude were zero, in the 
eikonal approximation that they used, the cross section would 
be independent of the sign of $a_I$. Since this ratio is 
small, but non-zero, this symmetry is only approximate and 
they found about the same large improvement in the agreement 
with data \cite{satta} for values of +10 and --15 (GeV/c)$^{-2}$ at 1.75 
GeV/c, per nucleon, +7.5 and --13 (GeV/c)$^{-2}$ at 1.25 
GeV/c per nucleon and +11.5 and --12 (GeV/c)$^{-2}$ at 1.08 
GeV/c per nucleon. Clearly the values from the phase shift 
analyses favor the negative sign although the trend with 
incident momentum seems to be contrary to that seen in Ref. 
\cite{fy}. However, except for the 1.75 GeV/A case, the 
values of $a_R$ are very different from those obtained from 
spin-averaged amplitudes (see Fig. \ref{solexpv} or 
\ref{solws1}). A second way to obtain the slope parameter is 
to fit the differential cross section with an exponential in 
$t$. In this case the spin-flip cross section is included 
which leads to a more nearly isotropic cross section and 
hence smaller values of $\ar$. Thus, with this choice some
effect of the spin flip is included in the representation.
The values for $\ar$ are believed to be equivalent at about 
600 MeV \cite{betasource}.

El-Gogary et al. \cite{elgogary3} also treated $\alpha-\alpha$ 
scattering and found (using different NN amplitude parameters) that 
a phase factor linear in the momentum squared of a value of +5 
(GeV/c)$^{-2}$ greatly improved the agreement with the data at 1.75 
GeV/c/A.

Usmani et al. \cite{usmani} considered the case of a modified 
helium wave function consisting of the sum of two Gaussian 
pieces.  They found that for $\alpha$ scattering on $^4$He only 
moderate corrections were seen. Since the data extend to a 
value of $q^2$ of 4 (GeV/c)$^2$ this may seem to be contrary 
to expectations. However, the $\alpha$$^4$He scattering is 
dominated by a large number of scatterings in this momentum 
transfer range so that a typical value of $q^2$ for one of 
the scatterings will be reduced by a factor equal to the 
number of scatterings, typically between 8 and 16. So the form 
factor needs to be accurate only up to a range of the order 
of $-t=0.25-0.5$\ (GeV/c)$^2$. 

\subsection{Heavier Nuclei}

Treating heavier nuclei, Lombard and Maillet \cite{lm} 
considered, in addition to the cross section, the asymmetry, 
$A$, and the spin rotation parameter, $Q$ so they were forced 
to include the spin dependent amplitude. While they used for 
the spin-independent amplitude a basic form with no phase 
variation, the basic form of the spin-flip amplitude included 
a phase variation consistent with phase shift analyses [see 
their formula (2)]. The phase variation in their basic 
spin-flip amplitude can be represented by a coefficient of $t$ 
of about 0.95 (GeV/c)$^{-2}$. They then introduced a  global 
phase variation which was applied to both amplitudes and 
estimated the phase parameter which would improve the 
agreement with the cross section and $A$ ($Q$ had not been 
measured at the time). They found that while $A$ was not very 
sensitive to a variation in this phase, the quantity $Q$ was. 
However, since there seems to be no reason to believe that the 
two amplitudes would have the same phase variation (or the 
difference they used), one can only conclude from their 
results that there is a sensitivity to the phase of the 
amplitudes.

\begin{figure}[htb]
\epsfig{file=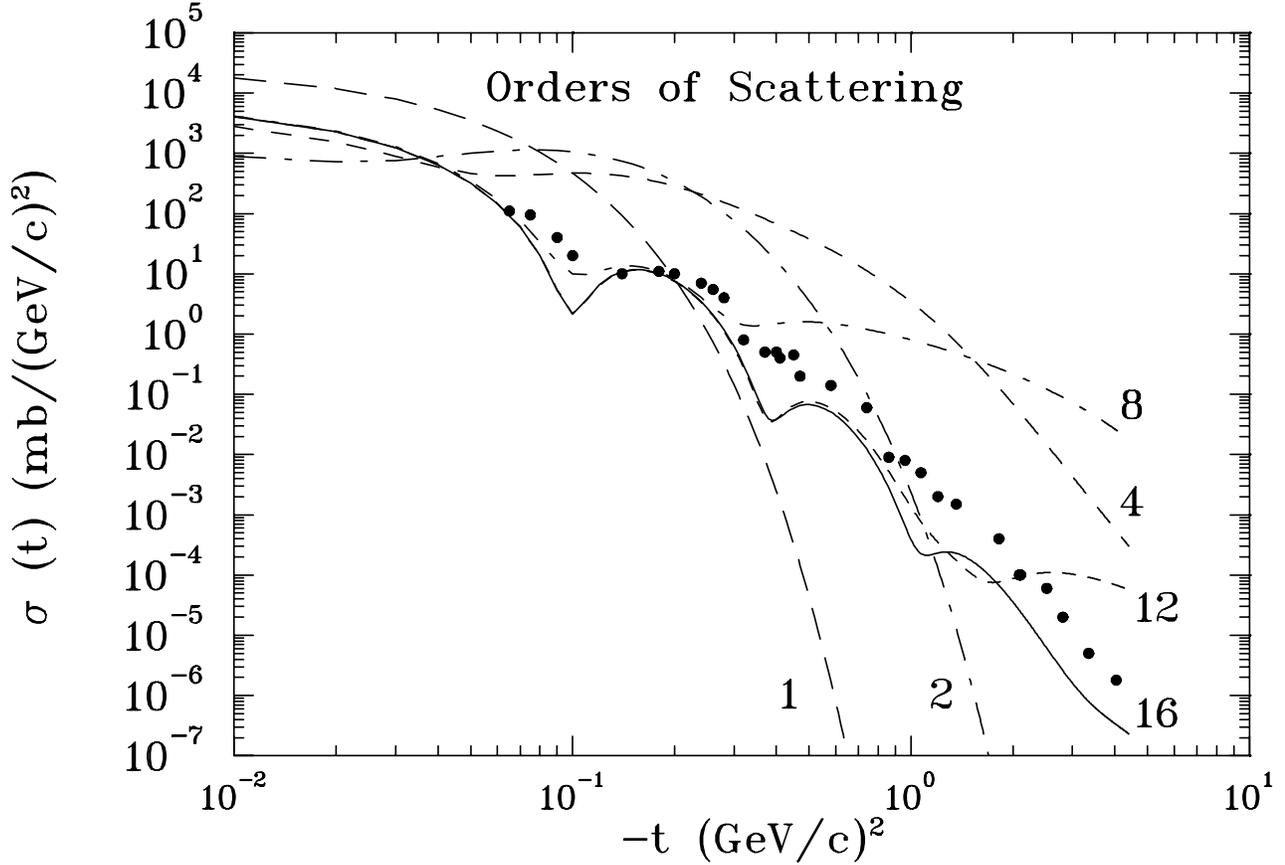,angle=90,height=4.5in}
\caption{Comparison of the cross section for $\alpha \alpha$ scattering 
calculated through various orders for $a_I=0$ and $\rho=-0.23$. The data 
are from Satta \ea \cite{satta}  }
\label{fy44order}
\end{figure}

\begin{figure}[htb] 
\epsfig{file=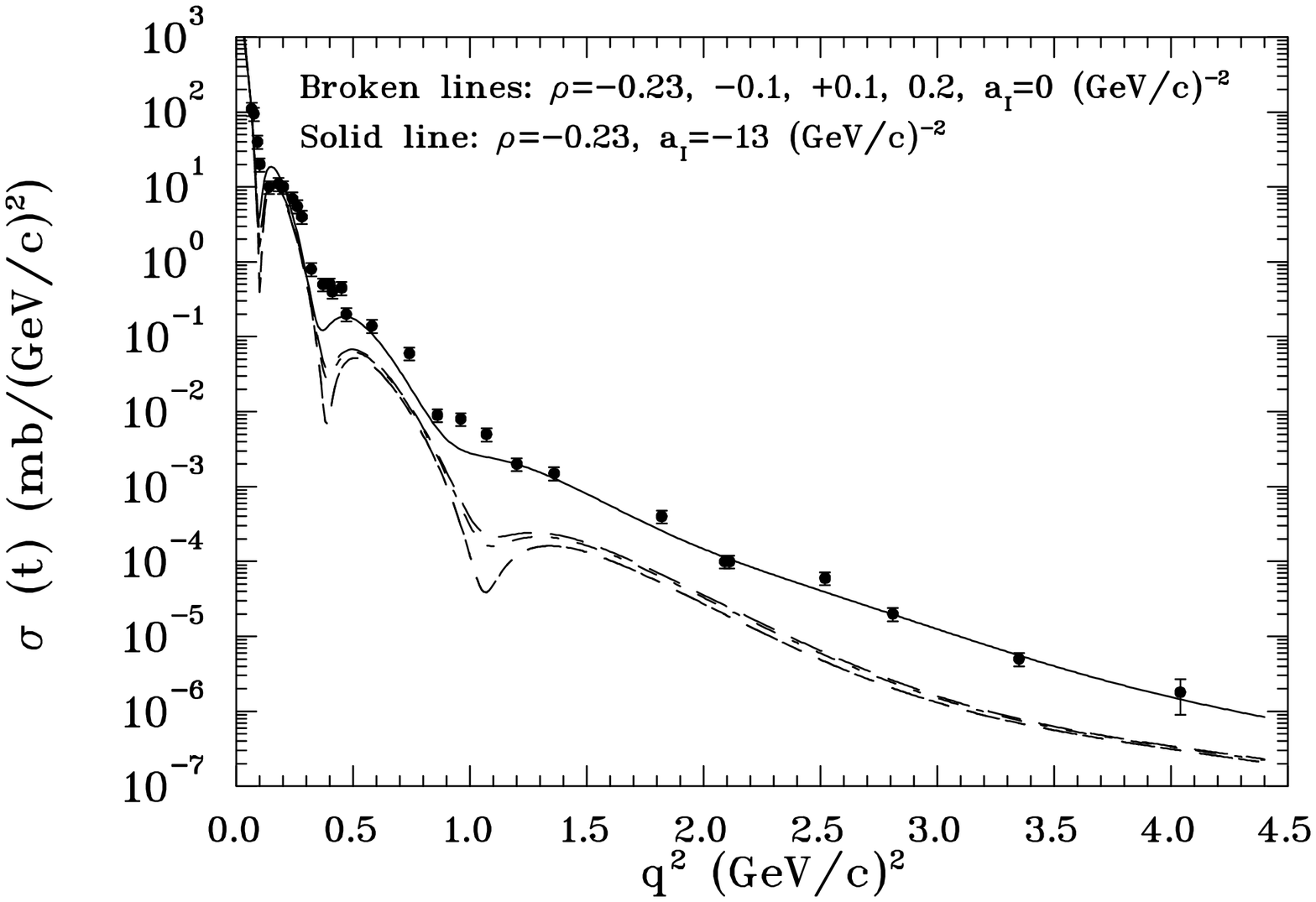,angle=90,height=3.in} 
\epsfig{file=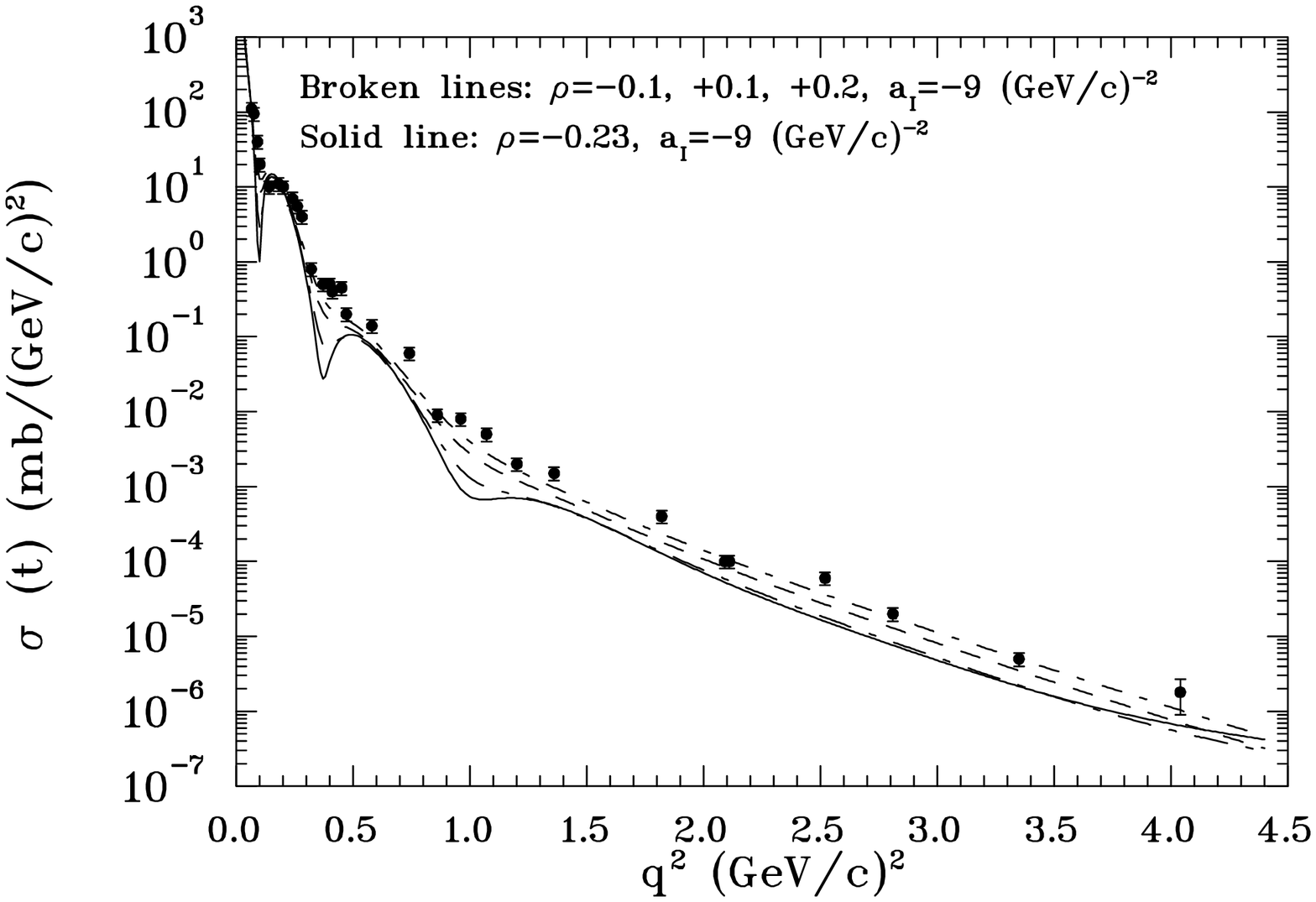,angle=90,height=3.in} 
\caption{Dependence of the cross section for $\alpha \alpha$\ scattering 
on $\rho$ and $a_I$. The lower panel shows that for $\rho=0.20$ the value 
of $\ai=-9$\ (GeV/c)$^{-2}$ gives an adequate fit to the data of Satta \ea 
\cite{satta}.} \label{fy44rho} 
\end{figure}

Lassaut, Lombard and Van de Wiele \cite{lassaut} considered 
an {\it additional} phase variation to be applied to 
representations of both amplitudes determined from amplitude 
analyses \cite{auger}. They also applied the same phase 
variation to both amplitudes. If one assumes that the 
non-spin-flip amplitude dominates the elastic cross section 
and adds the phase parameter they obtained (--0.25 fm$^2$=--6.4 
(GeV/c)$^{-2}$) to the one present in their basic amplitudes 
(--5.6 (GeV/c)$^{-2}$) one arrives at a value of (--12 
(GeV/c)$^{-2}$) for their incident momentum of 1.46 GeV/c, 
similar to the negative values --13 and --15 (GeV/c)$^{-2}$ 
obtained by Franco and Yin at nearby incident energies.

Auger and Lazard \cite{auger2} considered the effect of 
multiplying a global phase times a parameterization of both 
of the nucleon amplitudes.  Since they considered only 
asymmetry and spin rotation observables it is difficult to 
compare with their results. They considered phase factors 
with parameters close to those of Franco and Yin, however, 
since these must be added to the effective values already 
implicit in the phenomenological amplitudes they are not 
really comparable.

Chaumeaux \ea \cite{chaumeaux} also considered the addition of a phase 
variation in the scattering from heavier nuclei.

One expects that for all but the very lightest nuclei the sensitivity to 
such a phase will be small since the scattering at high energies is mainly 
determined by the radius and the diffuseness of the surface \cite{adl}. 
The influence of such a phase would mainly be contained in the region of 
the minima where many different multiple scattering and medium effects 
contribute.

\subsection{Extracting the Phase Variation from Multiple 
Scattering}

We have revisited the Glauber calculation of Franco and Yin \cite{fy} with 
a view to determine the effect of the possible nuclear modifications on 
the extraction of $a_I$ from the data \cite{satta} at 1.75 GeV/c/A beam 
momentum. For helium scattering from helium it is possible to do the full 
multiple scattering calculation in the eikonal approximation. An 
additional advantage is that the basic form factor of helium has no zero 
in the momentum range of interest whereas for heavier nuclei that is not 
true.  Thus, the minima are better understood.  They are not, however, 
simple interferences between different scattering orders (as they are 
approximately for proton scattering on $^4$He) but arise from a more 
complicated set of interferences.  Figure \ref{fy44order} shows the 
calculation with the amplitudes obtained from the sum up to a limited 
number of orders for $a_I=0$ and $\rho=-0.23$.  One sees, for example, 
that the first minimum is well defined in position only by summing 
through eighth order scattering and its depth is established by twelth 
order. One also sees that the first order scattering is negligible above 
about 0.6 (GeV/c)$^2$. For $\rho=\ai=0$, the individual orders of 
scattering are purely imaginary and alternate in sign. The effect of a 
finite value of $\rho$ is to multiply the $n$th order amplitude by 
$(1-i\rho)^n$.

One can be concerned that the effects of $\rho$ and $\ai$ may be confused. 
The result for the full multiple scattering calculation for 
$\alpha \alpha$ scattering is well below the data for large $q^2$ while in 
the case of the heavy nuclei, this is not in general true. There the major 
effect of setting $\ai$ to a non-zero value is seen in the minima.  
Figure \ref{fy44rho} shows results for variations in $\rho$ over the 
uncertainty observed in the NN data. The major effect for $\ai=0$ is 
to be seen only in the minima. For a finite value of $\ai$ of --13 
(GeV/c)$^{-2}$, this is not the case with the cross section away from the 
minima being increased as well.

\begin{figure}[htb] 
\epsfig{file=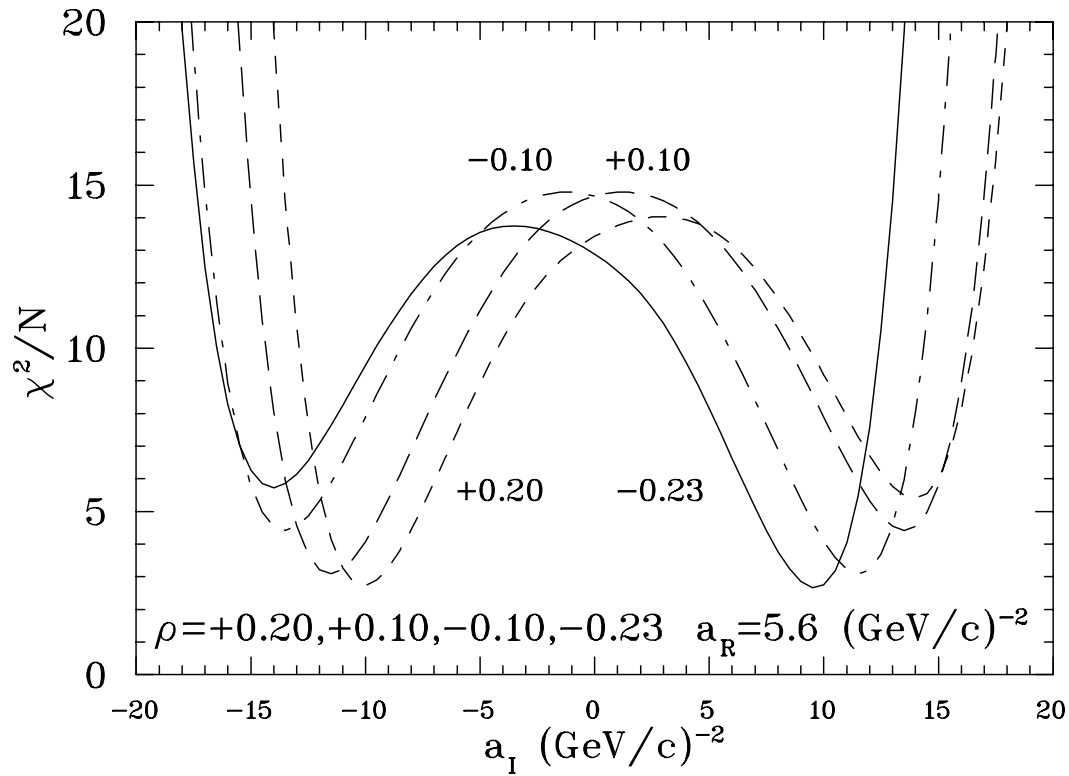,angle=90,height=4.in} 
\caption{Values of $\chi^2$ from a comparison with the measured cross 
section for $\alpha \alpha$\ scattering \cite{satta} 
as a function of $a_I$ for four different
values of $\rho$.  
} \label{fy44chirho5.6} 
\end{figure}

In an attempt to see how well the values of $\ai$ are determined from the 
$\alpha$ $^4$He data we have calculated a $\chi^2$ measure with regard to 
the data by Satta \ea\ \cite{satta}. Since no tabulated data were given we 
took the data from the plots including 20\% errors for most points. Figure 
\ref{fy44chirho5.6} shows $\chi^2/N$ for four selected values of $\rho$ 
where it is seen that there are two minima. If $\rho$ were zero, the 
$\chi^2/N$ curve would be symmetric.  For $\rho=+0.20$ the value of 
about --10 (GeV/c)$^{-2}$ is favored for the negative solution.

Looking at Figs.  \ref{solexpv} and \ref{solws1} it is seen that
these values of $\rho$ and $\ar$ might be possible by a combination
of increasing the radius of the real part of the interaction and
lowering the energy for the evaluation of the NN parameters 
although, even with these assumptions, one is at the very limit.

Before one can consider such a reconciliation of the multiple 
scattering and free space determinations of the phase variation
there are a number of corrections which must be treated.

a) The spin-flip may give substantial corrections through double
spin flip.

b) The off-shell corrections could give a contribution to the 
amplitude of similar nature to that of $\rho$ and $\ar$.

c) Short-range correlations in the $^4$He wave function will
modify the form factor.

d) Three-body forces could affect the scattering.

e) Standard corrections to Glauber theory \cite{eikonal} still
need to be considered.

This list is not meant to be exhaustive.

\section{Conclusions}

It has been shown that the phase of the strong scattering amplitude
is a sensitive function of the relative size of the radii of the real
and imaginary parts of a potential describing the interaction. This was
done with the consideration of three models of the potential.

We have found that the phase parameter is moderately well determined 
for the example case of K$^+$p scattering from standard amplitude 
analyses with the controlling principle being the unitary expansion 
of the amplitude. It is expected that the phase in the nucleon-nucleon
case would be better determined.

Since the effective amplitude in the nucleus can be taken as the free 
amplitude evaluated at a shifted (normally lower) energy one can 
perhaps understand the larger magnitude of the phase parameter than 
that seen in free space. Studies of this shift in the effective 
energy of the scattering amplitude have been made in the case of 
pion-nucleus scattering \cite{piscat} and the predictions of such an 
energy shift were verified \cite{cotting} experimentally. Similar
corrections have been calculated in this energy range for nucleon-nucleus
scattering \cite{mediump}.

This explanation may not be adequate to give a fit to the data so one 
is led to consider the possibility of a larger radius for the real 
part of the interaction due to a partial deconfinement in the nuclear 
medium. If this is true it may be a new way to study 
``non-classical'' modifications in the nuclear medium.

This work was supported by the National Science Foundation under 
Contract PHY-0099729.

\begin{appendix}

\section{Expansion of the NN Amplitude}

Starting from Eq. \ref{expand1} and using
\eq
\int_0^{\infty}bdbJ_0(bq)e^{-d^2b^2}=\frac{1}{2d^2}e^{-\frac{q^2}{4d^2}}
\qe
we can write
\eq
f(q)=-\frac{ikr_R^2}{2}\sum_{n=1}^{\infty} \frac{(-1)^n}{n!}
\sum_{m=0}^n \left( \begin{array}{c} n\\ m\end{array}\right)
\alpha^{n-m}\beta^m\frac{e^{-\frac{q^2r_R^2}{4(n-m+m\eta)}}}{n-m+m\eta}.
\label{40}\qe 

This equation is a generalization of Eq. 4.11 in Wallace \cite{eikonal}
for two independent radii. In the forward direction (q=0) we can write
\eq
f(0)=\frac{-ikr_R^2}{2}\sum_{n=1}^{\infty} \frac{(-1)^n}{n!}
\sum_{m=0}^n \left( \begin{array}{c} n\\ m\end{array}\right)
\alpha^{n-m}\beta^m\frac{1}{n-m+m\eta}
\qe
or
\eq
f(0)=\frac{-ikr_R^2}{2}\sum_{n=1}^{\infty} \frac{(-1)^n}{n!}
\int_0^{\infty}dt \sum_{m=0}^n \left( \begin{array}{c} n\\ 
m\end{array}\right)
\alpha^{n-m}e^{-(n-m)t}\beta^me^{-m\eta t}
\qe
\eq
=\frac{-ikr_R^2}{2}\int_0^{\infty}dt 
\sum_{n=1}^{\infty} \frac{(-1)^n}{n!}
\left(\alpha e^{-t}+\beta e^{-\eta t}\right)^n
=\frac{-ikr_R^2}{2}\int_0^{\infty}dt 
\left[e^{\left(-\alpha e^{-t}-\beta e^{-\eta t}\right)}-1\right].
\qe
Transforming the variable of integration from $t$ to $z$ with
 $z=e^{-\eta t}$ we have
\eq
f(0)=\frac{-ikr_R^2}{2\eta}\int_0^1dz\frac{
e^{-\alpha z^{\frac{1}{\eta}}}e^{-\beta z} -1}{z}.
\qe
Since $\alpha$ is thought of as smaller than $\beta$ (in absolute 
magnitude) we expand the exponential in $\alpha$ to find
\eq
f(0)=\frac{-ikr_I^2}{2}\int_0^1dz\frac{
\left[(1-\alpha z^{\frac{1}{\eta}}+\dots)e^{-\beta z} -1\right]}{z}
\qe
\eq
=\frac{-ikr_I^2}{2}\left\{\int_0^1dz\frac{\left[e^{-\beta z} -1\right]}{z}
+\sum_{\ell=1}^{\infty}\frac{(-\alpha)^{\ell}}{\ell !}\int_0^1 e^{-\beta 
z}z^{\frac{\ell}{\eta}
-1}\right\}. \label{46}
\qe

Since
\eq
\int_0^1dz\frac{\left[e^{-\beta z} -1\right]}{z}
=\sum_{n=1}^{\infty}\frac{(-\beta)^n}{n!n}
=-g_1-\ln (\beta)-E_1(\beta)
\qe
where $g_1$ is Euler's constant ($= 0.57721 \dots$) and $E_1(\beta)$ 
is the exponential integral, we have a closed form for the leading 
order which leads directly to Eq. \ref{totalcross} in the main text.

The full amplitude reads
\eq
f(q)=\frac{-ikr_R^2}{2}\sum_{j=0}^{\infty} \frac{1}{(j!)^2}
\left(\frac{-q^2r_R^2}{4}\right)^j \int_0^{\infty} t^j
\left[ e^{-\left(\alpha e^{-t}+\beta e^{-\eta t}\right)}-1\right]
dt
\qe
and upon using the same change of variable

\eq
f(q)=\frac{-ikr_I^2}{2}\sum_{j=0}^{\infty} \frac{1}{(j!)^2}
\left(\frac{q^2r_I^2}{4}\right)^j \int_0^1 dz (\ln z)^j
\frac{e^{-\beta z}e^{-\alpha z^{\frac{1}{\eta}}}-1}{z}.
\qe
Expanding the $\alpha$ exponential we have

\eq
f(q)=\frac{-ikr_I^2}{2}\sum_{j=0}^{\infty} \frac{1}{(j!)^2}
\left(\frac{q^2r_I^2}{4}\right)^j \int_0^1 dz (\ln z)^j
\left[\frac{e^{-\beta z}-1}{z}+\sum_{\ell=1}^{\infty}
\frac{(-\alpha)^{\ell}}{\ell !} z^{\frac{\ell}{\eta}-1}
e^{-\beta z}\right]
\qe
and since
\eq
\int_0^1 dz (\ln z)^j z^{\nu}=\frac{(-1)^j j!}{(\nu+1)^{j+1}}
\qe
we may finally write
\eq
f(q)=-\frac{kr_I^2}{2}\sum _{j=0}^{\infty}\frac{1}{j!}\left[
\sum_{n=1}^{\infty} \frac{(-\beta)^n}{n!n^{j+1}}+
\sum_{\ell=1}^{\infty}\frac{(-\alpha)^{\ell}}{\ell !}
\sum_{n=0}^{\infty} \frac{(-\beta)^n}{n!(\frac{\ell}{\eta}+n)^{j+1}}
\right]\left(\frac{-q^2r_I^2}{4}\right)^j. \label{a14}
\qe

\section{Properties of the function $\f$}

The power series definition of $\f(u,k,x)$ (\ref{lamdef})
is convergent for
all values of $u$ and $x$ but for large values of $x$ (greater
than about 40) it is numerically difficult to calculate in that
manner. We note that the relation
\eq
x\frac{d\f (0,k,x)}{dx}=\f (0,k-1,x)
\qe
follows from the series definition. From this equation and from the
known relation for $k=0$ and $k=1$
\eq
\f (0,0,x)=e^{-x}-1;\ \ \f (0,1,x)=-\ln(x)-g_1-E_1(x)
\qe
we can iterate to find the asymptotic behavior for large $x$.

The asymptotic behavior of $\f(0,k,x)$ is given by 
\eq
\f(0,k,x)\rightarrow-\sum_{i=0}^k\frac{ g_{k-i} [\ln(x)]^i}{i!}
\qe
for $k\ge 0$.
Note that $\f (0,k,x) \le 0$ and $\f (0,k,x)\ge \f (0,k+1,x)$ 
for $x \ge 0$ and all $k\ge 0$. 

The values of $\f(0,k,x)$ can be calculated to 5 significant figures with
the constants given in Table \ref{gs} for $x$ greater than 10. The
values for $x$ less than 40 can be easily calculated by the series.

For the variation in the first parameter, u, the relation
\eq
\f(u+1,k,x)=-\frac{d\f(u,k,x)}{dx}-\frac{1}{(u+1)^k}
\qe
allows the calculation the function for all values of u from those
between 0 and 1.
For small $u$ we can expand the sum to get
\eq
\f(u,k,x)=\f(0,k,x)-ku\f(0,k+1,x)+\dots
\qe

\begin{table}
$$ \begin{array}{|c|l|}
\hline
0&1.000000\\
\hline
1&0.577215664\\
\hline
2&0.9890560\\
\hline
3&0.9074791\\
\hline
4&0.9817279\\
\hline
5&0.9819955\\
\hline
\end{array}$$
\caption{Values of the coefficients $g_j$ determined numerically (except
for the first two).}\label{gs}
\end{table}

\end{appendix}

\end{document}